\documentclass[twocolumn]{aastex631}
\newcommand{\ysx}{\color{black}}
\usepackage{lineno}
\begin{document}
\title{Fast X-ray Transients as Premature-Dissipated
       Gamma-Ray Bursts with Magnetic-Dominated Jets}

\author{Shu-Xu Yi}
\affiliation{Institute of High Energy Physics, Chinese Academy of Sciences,
             Beijing 100049, China}
\author{Shao-Lin Xiong}
\affiliation{Institute of High Energy Physics, Chinese Academy of Sciences,
             Beijing 100049, China}
\shorttitle{FXT as Premature-Dissipated GRBs}
\shortauthors{S.-X. Yi, Shao-Lin Xiong}

\begin{abstract}
Fast X-ray transients (FXTs) are extragalactic X-ray flashes with little or no accompanying $\gamma$-ray emission. We propose that FXTs and classical gamma-ray bursts (GRBs) share a common origin in Poynting-flux-dominated jets powered by the Blandford--Znajek mechanism. In this picture, FXTs arise when the kink instability triggers premature dissipation at small radii ($\sim 10^{11}$~cm), yielding low Lorentz factors ($\Gamma \lesssim 50$) and high compactness ($\ell' \gg 1$) that thermalize the emission into the soft X-ray band. When the kink criterion is not met, the jet accelerates to ultra-relativistic velocities ($\Gamma > 100$) and dissipates at larger radii ($\sim 10^{15}$~cm) as a standard GRB. This scenario can naturally explain the observed thermal spectra, low $E_{\rm peak}$, supernova associations, jet-break absence, and day-scale afterglow delays across five well-studied FXTs. EP240801a is explained as a boundary event, where premature dissipation does not exhaust the full magnetic energy; the residual energy is further dissipated at a typical GRB radius, producing the observed $\gamma$-ray prompt emission with an $\sim 80$~s time lag. A key observational prediction is that FXTs may exhibit $e^+ e^-$ annihilation line features sweeping from a few MeV to hundreds of keV within $\sim 0.01$~s. 
\end{abstract}

\section{Introduction}
Fast X-ray transients (FXTs) were first identified as serendipitous X-ray flashes without detectable $\gamma$-ray emission in archival {\it Chandra} data \citep{jonker2013fxt,glennie2015fxt}, with systematic searches later revealing a population of extragalactic candidates \citep{quirola2022fxt}. The Einstein Probe (EP) has dramatically expanded this population. Among these, some have been observed with extensive multi-wavelength follow-ups, discovering FXTs with remarkably diverse properties: EP241113a at $z=1.53$ has an isotropic-equivalent energy $E_{\gamma,\rm iso}\sim 10^{51}$~erg comparable to classical gamma-ray bursts (GRBs), yet an exceptionally thermal soft spectrum with $E_{\rm p}\lesssim 1$~keV, deviating from the Amati relation at $>3\sigma$; its afterglow plateau implies a low bulk Lorentz factor $\Gamma\sim 20$, and the absence of a jet break to $\sim30$ days suggests a wide opening angle \citep{2026arXiv260326213D}. EP250108a at $z=0.176$ is even softer, with $E_{\rm iso}\lesssim 10^{49}$~erg and $E_{\rm peak}<1.8$~keV, at the extreme soft-and-weak end of the Amati relation, associated with the extremely luminous broad-lined Type Ic supernova (SN Ic-BL) SN~2025kg \citep{li2025ep250108a}. EP240414a at $z=0.40$ with $E_{X,\rm iso}\sim 5\times10^{49}$~erg lies below the Amati relation and exhibits a two-component light curve---early thermal emission followed by a non-thermal rebrightening peaking at $\sim3$ days---accompanying SN~2024gsa \citep{sun2025ep240414a,zheng2025ep240414a}. EP240801a at $z=1.67$ sits on the Amati relation with $E_{\rm p}=14.9$~keV and displays a two-stage prompt emission: a soft X-ray component lasting 148~s, followed by a harder gamma-ray component with an $\sim80$~s X-ray lead, consistent with a non-thermal cutoff power-law spectrum \citep{jiang2025ep240801a}. XRF~241001A at $z=0.57$ is a soft on-axis X-ray flash with $E_{\rm iso}\sim 8\times10^{49}$~erg and $E_{\rm peak}<10$~keV, associated with SN~2024aiiq (SN Ic-BL) \citep{schneider2026xrf241001a}.

\cite{dai2026ep241113a} proposed an NDAF-powered dirty fireball model in which a baryon-loaded jet with low terminal Lorentz factor produces thermal emission via a dissipative photosphere, naturally explaining EP241113a's soft thermal spectrum. The model is {\ysx essentially} non-magnetic \footnote{\ysx although the model allowed foor weak magnetization and the authors mentioned magnetic rreconnection as a possible additional heating resource.} and predicts {\ysx both} thermal {\ysx and non-thermal spectrum}\footnote{\ysx depends on the location of the dissipative
photosphere, it could produce a broadened, Comptonized spectrum in addition to the
thermal one \citep{giannios2012,beloborodov2013,beloborodov2017,samuelsson2023}.}. \cite{zheng2025ep240414a} proposed an off-axis jet--cocoon model in which early thermal emission arises from cocoon cooling and late non-thermal emission from the structured jet afterglow, successfully explaining EP240414a, yet the model's inner and outer cocoon components yield marginally discrepant progenitor radius estimates that may require additional heating. \citep{li2026magnetarpwn} proposed a magnetar-driven pair wind nebula model in which a rapidly rotating magnetar powers a structured jet and a PWN embedded in SN ejecta, reproducing the thermal-to-non-thermal-to-SN light curves of EP240414a and EP250108a, though the inferred PWN particle spectral indices deviate from the standard picture. \citep{srinivasaragavan2025ep250108a} proposed a collapsar jet choked in an extended circumstellar medium, explaining EP250108a's double-peaked optical light curve and the soft X-ray prompt emission. \citep{jiang2025ep240801a} considered three scenarios for EP240801a---an off-axis normal GRB, an intrinsically weak GRB jet, and a two-component structured jet---all of which fit the data but cannot be distinguished with current observations. The off-axis GRB interpretation is further challenged by \citep{chen2026ep}, who found that viewing-angle effects of standard Type II GRBs alone cannot account for the observed spectral peak energies of some energetic FXTs. \citep{schneider2026xrf241001a} interpreted XRF~241001A as an on-axis low-$\Gamma$ jet forming the low-energy tail of the long GRB population, though the classification method for this regime remains to be developed. While each model plausibly accounts for the event it was designed to explain, no single framework simultaneously addresses the thermal/non-thermal diversity and the three-order-of-magnitude span in isotropic energy across the FXT population.

The common SN Ic-BL association across all well-studied FXTs points to a collapsar central engine shared with classical GRBs. Two FXT events (EP240801a, EP250108a) lie on the Amati relation \citep{jiang2025ep240801a,li2025ep250108a}, and EP240414a's late afterglow requires the presence of a GRB-like relativistic jet \citep{zheng2025ep240414a}. In the GRB context, there is growing evidence for magnetically dominated jets as a natural outcome of the Blandford--Znajek mechanism \citep{blandford1977,tchekhovskoy2011}, with GRB~230307A providing direct evidence for minijet emission from a Poynting-flux-dominated outflow \citep{yi2025grb230307a}. Yet none of the existing FXT models incorporate a magnetically dominated jet as the primary energy carrier, despite magnetic jets being the population expectation for BZ-powered collapsars. This gap motivates us to explore whether FXTs and GRBs can be unified within a single magnetized jet framework.

In this paper, we propose that FXTs and GRBs share the same Poynting-flux-dominated jet engine, differing only in the dissipation radius---premature at the stellar surface for FXTs versus delayed to the saturation radius for GRBs---with the boundary set by a causal connectivity criterion for the kink instability. In Section~2 we elaborate the physical model, including jet acceleration, the kink criterion, and the compactness-based spectral classification. In Section~3 we compare the model with the features of five observed FXTs. In Section~4 we discuss the implications, predictions, and limitations of the model. 

We note that contemporaneous soft X-ray counterparts of GRBs---e.g., EP250704a/GRB 250704B \citep{li2026ep250704a}, and GRB 230307A detected by EP's pathfinder, LEIA \citep{sun2025grb230307a}---are also sometimes referred to as FXTs observationally. These have non-thermal spectra that can be naturally explained as the low-energy extension of the GRB prompt emission, and are not included in the discussion of this paper.

\begin{figure*}
    \centering
    \includegraphics[width=\linewidth]{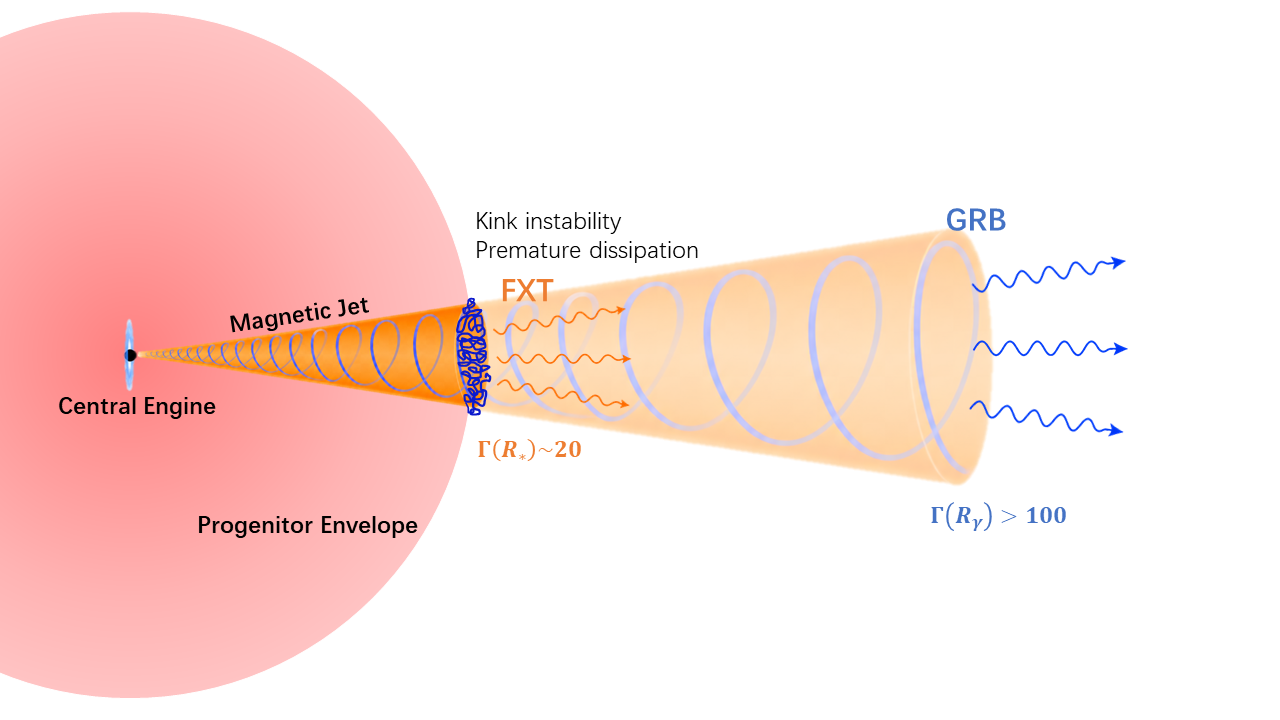}
    \caption{Schematic of the jet evolution scenario. A Poynting-flux-dominated jet is launched from the central engine and propagates through the progenitor envelope. Kink instability may develops in the magnetically dominated flow, triggering premature magnetic dissipation at low Lorentz factor, producing the fast X-ray transient (FXT). In the sketch, we also plot the case when the premature dissipation is absence, the jet continues accelerating to ultra-relativistic Lorentz factor and produces a classical gamma-ray burst (GRB) via dissipation at larger radii.}
    \label{fig:ill}
\end{figure*}
\section{The model}
We consider a Poynting-flux-dominated relativistic jet launched by the Blandford--Znajek mechanism during the core collapse of a massive progenitor star. At the jet base, the energy flux is carried primarily by magnetic Poynting flux, characterized by the initial magnetization $\sigma_0 \equiv \dot{E}_{\rm Poynting}/\dot{E}_{\rm kinetic} \gg 1$. The jet accelerates via the ideal-MHD magnetic pressure gradient $\nabla(B^2/8\pi)$, following $\Gamma(R) \propto R^{1/3}$ at the acceleration phase. 

The Lorentz factor at the stellar surface $R_*$ is therefore
\begin{equation}
\Gamma(R_*) = \Gamma_\infty \left(\frac{R_*}{R_{\rm sat}}\right)^{1/3}
           = \sigma_0^{3/2} \left(\frac{R_*}{R_{\rm sat}}\right)^{1/3}.
\label{eq:gRstar}
\end{equation}
For a typical GRB, the jet continues accelerating until it reaches the saturation radius $R_{\rm sat}$; thereafter it coasts at constant $\Gamma_\infty$ until, at some radius $R_\gamma$, magnetic energy is dissipated into radiation, producing the prompt gamma-ray emission.

Dissipation can also occur much earlier, at radii well below $R_{\rm sat}$, if the type II kink instability \citep{bromberg2016} triggers global magnetic reconnection across the jet cross-section. The scenario is sketched in figure \ref{fig:ill}. In a relativistic jet shell, two patches separated by an angular distance $\Delta\theta > 1/\Gamma$ are causally disconnected from each other \citep{2012MNRAS.421..570G,2025JHEAp..4700359Y,2026JHEAp..5300607Y}. A coherent global kink can therefore develop only if the entire jet cross-section lies within a single causally connected patch---that is, if the jet opening angle $\theta_{\rm J}$ satisfies $\theta_{\rm J}\Gamma(R) < 1$ at the radius $R$ where the kink mode grows, triggering magnetic reconnection that dissipates the stored Poynting flux; we refer to this as premature dissipation. Conversely, if $\theta_{\rm J}\Gamma(R) > 1$, the local kink mode is suppressed and never develops into a global dissipation event. We evaluate the premature dissipation condition at the stellar breakout radius $R_*$, since this is the minimal radius at which dissipation can produce an observable signal. If the condition is not met at $R_*$, it will not be met at any $R>R_*$ because $\Gamma$ continues to grow, making $\theta_{\rm J}\Gamma$ even larger. On the other hand, the simple kink criterion $\theta_J\Gamma<1$ is in fact more easily satisfied below $R_*$. The reason $R_*$ is treated as special is that the pressure-balance condition changes suddenly at that location; a mechanism demonstrating that this increases the likelihood of kink occurrence is yet to be established by future theoretical and numerical studies. If dissipation occurs deep below $R_*$, it may be trapped and might eventually escape with the outflow; We expect such emission to be inconsistent with the observed FXT, in that its timescale would be much longer and its spectrum would soften to the ultraviolet to optical range. Therefore, within the framework of our model we treat $R_*$ as the first radius at which dissipation produces an observable signature, rather than as a full justified dynamically preferred site.

\begin{figure*}
    \centering
    \includegraphics[width=\linewidth]{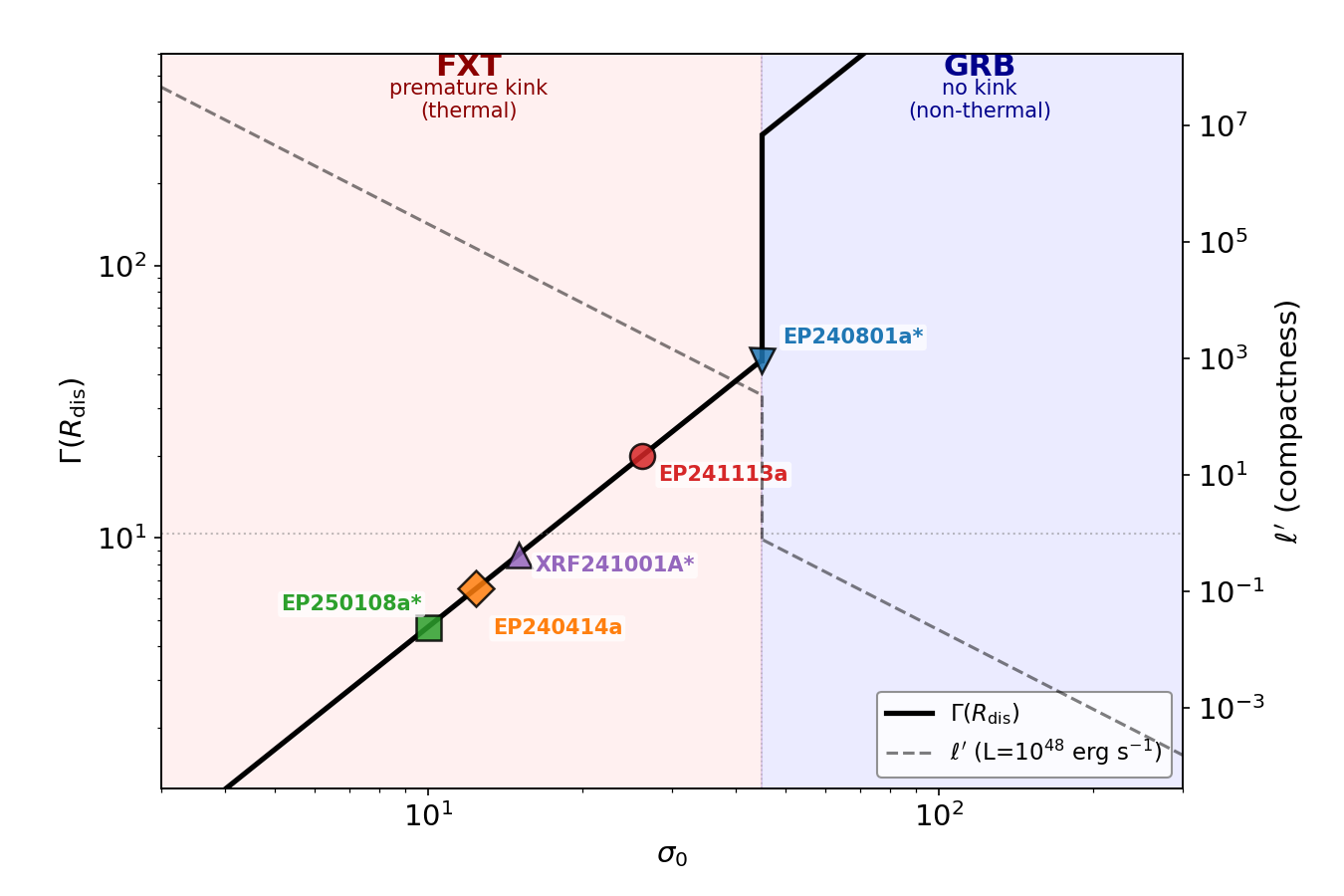}
    \caption{
  Model prediction for $\Gamma(R_{\rm dis})$ (solid black, left axis)
  and $\ell'$ (dashed, right axis, for $L=10^{48}$~erg~s$^{-1}$)
  as functions of $\sigma_0$.
  Red shading: FXT regime ($\sigma_0 < 45$, premature dissipation at $R_*$).
  Blue shading: GRB regime ($\sigma_0 > 45$, dissipation beyond $R_*$).
  The vertical segment at $\sigma_0 = 45$ connects the two branches.
  Observed events are marked at their inferred $\sigma_0$:
  EP241113a ($\Gamma \approx 20$; Dai et al.\ 2026)
  and EP240414a ($\Gamma \approx 6.5$, from the inner cocoon constraint of
  Zheng et al.\ 2025) have $\Gamma$ measurements in the literature;
  events marked with $^{*}$ (EP250108a, XRF\,241001A) lack a precise
  $\Gamma$ measurement and are shown at illustrative $\sigma_0$ values.
  EP240801a$^{*}$ is placed at the FXT/GRB boundary ($\sigma_0 = 45$)
  by model construction; its $\Gamma$ is not independently constrained
  by observations.}
    \label{fig:params}
\end{figure*}

Based on the above argument, the dissipation of magnetic energy in the jet can occur at two distinct radii: at the kink-induced premature location $R_{\rm pre} \sim R_*$, or beyond the saturation radius at $R_{\gamma} \sim 10^{15}$~cm, as in ordinary GRBs.
Adopting fiducial parameters $R_* = 10^{11}$~cm (typical Wolf--Rayet star), $\theta_{\rm J} = 1.3^\circ$, and $R_{\rm sat} = 3 \times 10^{13}$~cm (canonical GRB saturation radius), we show the resulting two-branch behavior of $\Gamma$ at the dissipation radius in Figure~\ref{fig:params}.
As the figure illustrates, the bifurcation is controlled by a single parameter $\sigma_0$.
Below a critical value $\sigma_0 \approx 45$, premature dissipation occurs at a small radius, yielding $\Gamma(R_{\rm dis}) \lesssim 50$, as implied by the recent FXT observations; above this threshold, premature dissipation does not occur, and the jet accelerates to $\Gamma \sim 100$--$300$ before dissipating, as in ordinary GRBs.

The premature dissipation at a smaller radius not only yields a lower Lorentz factor, but also places the emission in the compactness regime.
The comoving compactness parameter is defined as
\begin{equation}
\ell' = \frac{L \sigma_T}{4\pi \, R_{\rm dis} \, m_e c^3 \, \Gamma(R_{\rm dis})^3},
\label{eq:compactness}
\end{equation}
where $L$ is the isotropic-equivalent luminosity and $\sigma_T$ the Thomson cross-section \citep{peer2004}.
When $\ell' \gg 1$, photons efficiently pair-produce, rendering the emission region optically thick and thermalizing the radiation; the resulting spectrum is thermal, as observed in FXTs.
When $\ell' \ll 1$, the emission is optically thin and non-thermal, as in classical GRBs.
In Figure~\ref{fig:params} we also plot $\ell'$ as a function of $\sigma_0$.
As the figure shows, $\ell'$ exhibits the same bifurcation behavior: for the FXT branch ($\sigma_0 < 45$, $R_{\rm dis} \sim R_*$), the small radius and low $\Gamma$ combine to give $\ell' \gg 1$; for the GRB branch ($\sigma_0 > 45$, $R_{\rm dis} \sim R_{\rm sat}$), the much larger dissipation radius suppresses $\ell'$ below unity.

When $\ell' \gg 1$, the dissipation zone is a pair fireball, composed of photons and electron/positron pairs. The compactness condition in equation (\ref{eq:compactness}) assumes that all of the photon energy can participate in electron--positron pair production. The prerequisite for this is that the centre-of-mass energy of two photons exceeds the electron rest-mass threshold. When magnetic reconnection just occurs, most of the initial photons can satisfy the condition, so the compactness condition is physically meaningful. But subsequently the photons are rapidly thermalised, and the mean temperature becomes $\langle E^\prime \rangle \sim 2.7\,k_{\rm B}T'$. For a radiation dominated flow, $u'=aT'^4$ with $u'=L_{\rm iso}/(4\pi R_*^2\Gamma^2c)$, so that 
\begin{equation} 
\langle E^\prime\rangle \sim 2.7k_{\rm B}\left(\frac{L_{\rm iso}}{4\pi R_*^2\Gamma^2ac}\right)^{1/4}. \end{equation} 
Adopting the typical values $L_{\rm iso}=10^{48}~\mathrm{erg~s^{-1}}$, $R_*=10^{11}$~cm and $\Gamma\sim20$, we find that the mean energy of the photons after thermalisation is only $\sim 0.7$~keV, so that these photons can hardly participate in further pair production and escape in situ, emerging as observed soft X-ray photon with thermal spectrum. This physical process is the same as that described in \citet{ruffini1999pair}, \citet{ruffini2000pair}, \citet{bianco2001spike} and \citet{yi2024mevline}.

The above mentioned thermalisation happened in a timescale $\tau^\prime_{\rm th}$, which can be estimated as follows: 
\begin{equation}
    \tau^\prime_{\rm th}\sim N_{\rm th}\lambda /c,
\end{equation}
where $N_{\rm th}$ is the number of scattering for a high energy photon to be thermalised; For a photon with initial energy $E^\prime_i\sim$MeV, the number of scattering needed for it to be thermalised to $E^\prime_f\sim$keV is \citep{pozdnyakov1983}:

\begin{equation} 
N\sim\frac{\ln(E'_i/E'_f)}{E'_f/m_ec^2}\sim 10^3
\end{equation}
and $\lambda$ is the mean-free-path of the photon, which is $\lambda=\Delta R'/\ell'$. {\ysx Here we adopt the assumption of \citet{peer2004} that the comoving width of the dissipation shell is $\Delta R' = R_{\rm dis}/\Gamma$; under this assumption $\ell'$ is equivalent to the Thomson optical depth of the shell.} Therefore, the thermalisation timescale in the central engine frame is:

\begin{equation}
\tau_{\rm th}\sim\Gamma\tau^\prime_{\rm th}\sim\frac{N_{\rm th}R_*}{\ell'c}.
\end{equation}

For $\ell'\gg N_{\rm th}$, this timescale is much shorter than $R_*/c$, and the jet advances only a small fraction of $R_*$ before the radiation thermalises, so that the photons can be regarded as escaping essentially at $R_*$. For $\ell'\sim N_{\rm th}$, however, the thermalisation timescale is comparable to the expansion timescale, and the escaping radius $R_{\rm ph}$ exceeds $R_*$ but remains of the same order; in this regime the spectrum may not be fully thermalised. The minimum variability timescale of a single emission episode is then set by the angular spreading across the jet, $\Delta t_{\rm var} \simeq (1+z)\,R_{\rm ph}\theta_J^2/(2c) \sim 10^{-3}$~s (the jet opening angle, rather than the $1/\Gamma$ cone, since $\theta_J\Gamma<1$). This provides a testable prediction of the MVT of FXTs, which is currently not measured because the small number of counts does not allow a time-bin analysis on such short timescales.

\section{Confronting observations}
The low Lorentz factor and high compactness at premature dissipation naturally explain the observed low $E_{\rm peak}$ and thermal spectra of the FXT emission in EP241113a, EP240414a, EP250108a, XRF241001A, and EP240801a. The association of Type Ic supernovae with EP240414a, EP250108a, and XRF241001A is also consistent with the core-collapse progenitors of our model. Premature dissipation consumes a large fraction of the stored magnetic energy in the jet; consequently, for most FXTs, further gamma-ray emission cannot occur (the boundary event EP240801a is discussed separately below). The residual {\ysx magnetic} energy of the jet can still be deposited into the ambient medium at larger radii, producing the observed afterglows in EP241113a and EP240414a ({\ysx See discussion}). Dai et al.\ (2026) infer a wide opening angle for EP241113a from the absence of a jet break out to $\sim30$\,days. In the standard scenario, a jet break occurs when $\theta_J\Gamma$ transits from $>1$ to $<1$ during the deceleration phase. In our model, by contrast, $\theta_J\Gamma < 1$ holds at all times, so no jet break is expected regardless of the opening angle. For EP240414a, the $\sim3$\,day delay between the prompt FXT and the afterglow peak has been attributed to an off-axis core jet with a large Lorentz factor ($\Gamma \sim 300$; Zheng et al.\ 2025). In our scenario, the low Lorentz factor ($\Gamma \lesssim 10$) inferred from the prompt FXT naturally yields a deceleration timescale $t_{\rm dec}$ of order days, without requiring a large off-axis viewing angle.

EP240801a exhibits a two-stage prompt emission \citep{jiang2025ep240801a}: a soft X-ray component lasting $\sim$148\,s, followed $\sim$80\,s later by a harder $\gamma$-ray component.  In our picture, this burst may sit at the boundary between the two branches.  A kink develops marginally at $R_*$, triggering reconnection that produces thermal soft X-rays at high compactness ($\ell' \approx 10^3 \gg 1$).  This dissipation destroys the large-scale ordered field that drives acceleration, causing the jet to coast at $\Gamma \sim 50-100$ thereafter. The jet then coasts to a second dissipation radius $R_\gamma$, where a separate energy release---via internal shocks or reconnection of residual fields---produces the non-thermal $\gamma$-ray pulse. The relative low $E_{\rm peak}$ of the $\gamma$-ray prompt emission consistent with the low Lorentz factor.  At $R_\gamma$, the compactness drops below unity ($\ell' \approx 0.1$), corresponding to the non-thermal spectrum. The observed time-lag between these two stages can be estimated via: 

\begin{equation}
    \Delta t_{\rm obs} = (1+z)\frac{R_{\rm\gamma}}{2c\Gamma^2}
\end{equation}

Inserting a typical $R_\gamma=4\times10^{15}$\,cm, and $\Gamma=50$ results in $\Delta t$=71.2\,s, which is in nice agreement with the observed 80\,s time lag and shows the self-consistency of the scenario.

\section{Discussion}
A distinguishing prediction of our model is the potential of $e^+e^-$ annihilation features in the prompt emission of FXTs. Premature dissipation at $R_*$ occurs in the highly compact regime ($\ell' \sim 10^3$--$10^4$), where efficient $\gamma\gamma$ pair production drives the photon-pair fluid into thermal equilibrium. Since the thermalised photons do not have enough energy to participate the further pair creation and escaped at a radius $R_{\rm ph}\sim R_{\rm diss}$, a residual population of pairs annihilate essentially at $R_{\rm ph}$, and produces a narrow $511$~keV line in the comoving frame, Doppler-boosted. The above scenario is analogous to the annihilation features seen in GRB~221009A \citep{yi2024mevline}. Owing to the high-latitude curvature effect \citep{zhang2024grb221009a}, the line energy sweeps downward as $E_{\rm line} \propto (t-t_0)^{-1}$, with the timescale proportional to $R_{\rm line}$. In GRB~221009A, the annihilation features swept down from $37$ to $6$~MeV over $\sim135$~s, with estimated $R_{\rm line} \sim 10^{16}$~cm and $\Gamma\sim400$ \citep{2024SCPMA..6789511Z,yi2024mevline,zhang2024grb221009a}. Scaling down to lower Lorentz factor and smaller $R_{\rm line}$, the annihilation features in FXTs are therefore expected to sweep from a few MeV to hundreds of keV within $\sim10^{-2}$~s. A templated photon-by-photon search in gamma-ray monitors such as GECAM or {\it Fermi} for the sweeping line feature is therefore recommended. 

{\ysx Unlike the classical afterglow model driven by the baryonic kinetic energy, the afterglow in this framework arises from the residual Poynting flux, which is transferred to the shocked circumburst medium, where it is dissipated and radiated \citep{kennel1984,zhang2005,giannios2008,mimica2010}. This requires that the residual magnetization after the first, premature dissipation still satisfies $\sigma>1$. The prompt FXT emission and the afterglow then jointly constrain the total jet energy and their ratio. Since the two luminosities are of the same order of magnitude, the fraction of the magnetic energy dissipated at $R_{\rm dis}$ is of order unity, and the framework is self-consistent.}

In our framework, two-stage events like EP240801a sit at the boundary
of the bifurcation.  A natural expectation is therefore that such
events are rare in the FXT population.  Furthermore, their soft X-ray
emission is expected to have a higher $E_{\rm peak}$ among FXTs, while
their $\gamma$-ray $E_{\rm peak}$ should be lower than that of typical
GRBs. This prediction can be tested against future observations with a larger FXT sample. 

Our model is currently a toy model with several simplifications: a top-hat jet with constant $\theta_{\rm J}$ is assumed, whereas magnetic collimation and lateral expansion may cause the opening angle to evolve; the kink criterion is a causal-connectivity argument rather than a self-consistent simulation; detailed afterglow fitting in the low-$\Gamma$ regime has not been performed; and the self-consistent picture presented here does not preclude additional radiative components from other mechanisms (e.g., nested cocoon jet) in individual events. The dirty fireball scenario, on the other hand, is a exclusively different branch from this model, and cannot coexist in the same burst. 

Despite these above mentioned simplifications, the model provides a unified framework in which FXTs and GRBs originate from the same magnetized jet engine, with a single parameter $\sigma_0$ governing the key observational properties---low $E_{\rm peak}$, thermal spectra, and the absence of $\gamma$-ray prompt emission for most sources while retaining afterglow emission. Yet the three-order-of-magnitude span in isotropic-equivalent energy is not controlled by the diversity of $\sigma_0$, but by other independent parameters such as the engine luminosity and the duration of the central engine. Future, more detailed modeling would address the issues enumerated above.
\section{Acknowledgement}
We thank Prof. Gao He for useful discussions. This work was supported by the China’s Space Origins Exploration Program and Institute of High Energy Physics under the fund number E25155U110. The authors acknowledge the use of DeepSeek-V4-Pro for language editing and plotting codes generation.  All content generated by the LLM was reviewed and verified by the authors.

\bibliography{draft_v3}

@ARTICLE{2026arXiv260326213D,
       author = {{Dai}, C.-Y. and {Quirola-V{\'a}squez}, J. and {Wang}, Y.-H. and {Li}, H.-L. and {Yang}, J. and {Chen}, X.-L. and {Wang}, A.-L. and {Sun}, H. and {Wang}, X.-Y. and {Zhang}, B. and {Jonker}, P.~G. and {Liu}, Y. and {Yuan}, W. and {Xu}, D. and {Dai}, Z.-G. and {Ravasio}, M.~E. and {Piro}, L. and {O'Brien}, P. and {Stern}, D. and {Zhang}, H.-M. and {Yang}, Y.-P. and {An}, T. and {Qiu}, Y.-L. and {Xin}, L.-P. and {Li}, W.-X. and {Liu}, R.-Y. and {Wu}, X.-F. and {Wang}, C.-Y. and {Wei}, D.-M. and {Huang}, Y.-F. and {Bauer}, F.~E. and {Lei}, W.-H. and {Zhang}, B.-B. and {Sun}, N.-C. and {Gao}, H. and {Dhillon}, V.~S. and {An}, J. and {Bai}, C.-H. and {Martin-Carrillo}, A. and {Cheng}, H.-Q. and {Chacon Chavez}, J.~A. and {Chen}, Y. and {Du}, G.-W. and {van Dalen}, J.~N.~D. and {Esamdin}, A. and {Fan}, Y.-Z. and {Gao}, X. and {Harrison}, F. and {Hu}, J.-W. and {Huang}, M.-Q. and {Jia}, S.-M. and {Levan}, A.~J. and {Li}, C.-K. and {Li}, D.-Y. and {Liang}, E.-W. and {Littlefair}, S. and {Liu}, X.-W. and {Liu}, Z.-Y. and {Ling}, Z.-X. and {Malesani}, D.~B. and {Pan}, H.-W. and {Rodriguez}, A. and {Rossi}, A. and {Mata S{\'a}nchez}, D. and {S{\'a}nchez-Sierras}, J. and {Sun}, X.-J. and {Torres}, M.~A.~P. and {van Hoof}, A.~P.~C. and {Wang}, X.-F. and {Wu}, Q.-Y. and {Xu}, X.-P. and {Xu}, Y.-F. and {Yu}, Y.-W. and {Zhang}, C. and {Zhang}, M.-H. and {Zhang}, S.-N. and {Zhang}, Y. and {Zhang}, Y.-H. and {Zhu}, Z.-P.},
        title = "{An energetic dirty fireball detected in soft X-rays}",
      journal = {arXiv e-prints},
         year = 2026,
        month = mar,
          eid = {arXiv:2603.26213},
        pages = {arXiv:2603.26213},
          doi = {10.48550/arXiv.2603.26213},
archivePrefix = {arXiv},
       eprint = {2603.26213},
 primaryClass = {astro-ph.HE},
       adsurl = {https://ui.adsabs.harvard.edu/abs/2026arXiv260326213D}
}

@ARTICLE{dai2026ep241113a,
       author = {{Dai}, Cui-Yuan and {Wang}, Xiang-Yu and {Zhang}, Bing},
        title = "{EP241113a: Dissipative Photospheric Emission from a Dirty Fireball}",
      journal = {\apjl},
         year = 2026,
        month = aug,
       volume = {1006},
       number = {2},
          eid = {L62},
        pages = {L62},
          doi = {10.3847/2041-8213/ae8ba8},
archivePrefix = {arXiv},
       eprint = {2608.03043},
 primaryClass = {astro-ph.HE},
       adsurl = {https://ui.adsabs.harvard.edu/abs/2026ApJ..1006L..62D}
}

@ARTICLE{li2025ep250108a,
   author = {{Li}, W.-X. and {Zhu}, Z.-P. and {Zou}, X.-Z. and {Geng}, J.-J. and {Liu}, L.-D. and {Wang}, Y.-H. and others},
   title = "{An extremely soft and weak fast X-ray transient associated with a luminous supernova}",
   journal = {arXiv e-prints},
   year = 2025,
   eprint = {2504.17034},
   archivePrefix = {arXiv},
}

@ARTICLE{sun2025ep240414a,
       author = {{Sun}, H. and {Li}, W.-X. and {Liu}, L.-D. and {Gao}, H. and {Wang}, X.-F. and {Yuan}, W. and {Zhang}, B. and {Filippenko}, A.~V. and {Xu}, D. and {An}, T. and {Ai}, S. and {Brink}, T.~G. and {Liu}, Y. and {Liu}, Y.-Q. and {Wang}, C.-Y. and {Wu}, Q.-Y. and {Wu}, X.-F. and {Yang}, Y. and {Zhang}, B.-B. and {Zheng}, W.-K. and {Ahumada}, T. and {Dai}, Z.-G. and {Delaunay}, J. and {Elias-Rosa}, N. and {Benetti}, S. and {Fu}, S.-Y. and {Howell}, D.~A. and {Huang}, Y.-F. and {Kasliwal}, M.~M. and {Karambelkar}, V. and {Stein}, R. and {Lei}, W.-H. and {Lian}, T.-Y. and {Peng}, Z.-K. and {Frederiks}, D.~D. and {Ridnaia}, A.~V. and {Svinkin}, D.~S. and {Wang}, X.-Y. and {Wang}, A.-L. and {Wei}, D.-M. and {An}, J. and {Andrews}, M. and {Bai}, J.-M. and {Dai}, C.-Y. and {Ehgamberdiev}, S.~A. and {Fan}, Z. and {Farah}, J. and {Feng}, H.-C. and {Fynbo}, J.~P.~U. and {Guo}, W.-J. and {Guo}, Z. and {Hu}, M.-K. and {Hu}, J.-W. and {Jiang}, S.-Q. and {Jin}, J.-J. and {Li}, A. and {Li}, J.-D. and {Li}, R.-Z. and {Liang}, Y.-F. and {Ling}, Z.-X. and {Liu}, X. and {Mao}, J.-R. and {McCully}, C. and {Mirzaqulov}, D. and {Newsome}, M. and {Padilla Gonzalez}, E. and {Pan}, X. and {Terreran}, G. and {Tinyanont}, S. and {Wang}, B.-T. and {Wang}, L.-Z. and {Wen}, X.-D. and {Xiang}, D.-F. and {Xue}, S.-J. and {Yang}, J. and {Zhu}, Z.-P. and {Cai}, Z.-M. and {Castro-Tirado}, A.~J. and {Chen}, F.-S. and {Chen}, H.-L. and {Chen}, T.-X. and {Chen}, W. and {Chen}, Y.-H. and {Chen}, Y.-F. and {Chen}, Y. and {Cheng}, H.-Q. and {Cordier}, B. and {Cui}, C.-Z. and {Cui}, W.-W. and {Dai}, Y.-F. and {Fan}, D.-W. and {Feng}, H. and {Guan}, J. and {Han}, D.-W. and {Hou}, D.-J. and {Hu}, H.-B. and {Huang}, M.-H. and {Huo}, J. and {Jia}, S.-M. and {Jia}, Z.-Q. and {Jiang}, B.-W. and {Jin}, C.-C. and {Jin}, G. and {Kuulkers}, E. and {Li}, C.-K. and {Li}, D.-Y. and {Li}, J.-F. and {Li}, L.-H. and {Li}, M.-S. and {Li}, W. and {Li}, Z.-D. and {Liu}, C.-Z. and {Liu}, H.-Y. and {Liu}, H.-Q. and {Liu}, M.-J. and {Lu}, F.-J. and {Luo}, L.-D. and {Ma}, J. and {Mao}, X. and {Nandra}, K. and {O'Brien}, P. and {Pan}, H.-W. and {Rau}, A. and {Rea}, N. and {Sanders}, J. and {Song}, L.-M. and {Sun}, S.-L. and {Sun}, X.-J. and {Tan}, Y.-Y. and {Tang}, Q.-J. and {Tao}, Y.-H. and {Wang}, H. and {Wang}, J. and {Wang}, L. and {Wang}, W.-X. and {Wang}, Y.-L. and {Wang}, Y.-S. and {Xiong}, D.-R. and {Xu}, H.-T. and {Xu}, J.-J. and {Xu}, X.-P. and {Xu}, Y.-F. and {Xu}, Z. and {Xue}, C.-B. and {Xue}, Y.-L. and {Yan}, A.-L. and {Yang}, H.-N. and {Yang}, X.-T. and {Yang}, Y.-J. and {Zhang}, C. and {Zhang}, J. and {Zhang}, M. and {Zhang}, S.-N. and {Zhang}, W.-D. and {Zhang}, W.-J. and {Zhang}, Y.-H. and {Zhang}, Z. and {Zhang}, Z. and {Zhang}, Z.-L. and {Zhao}, D.-H. and {Zhao}, H.-S. and {Zhao}, X.-F. and {Zhao}, Z.-J. and {Zhou}, Y.-L. and {Zhu}, Y.-X. and {Zhu}, Z.-C. and {Zou}, H.},
        title = "{A fast X-ray transient from a weak relativistic jet associated with a type Ic-BL supernova}",
      journal = {Nature Astronomy},
         year = 2025,
        month = jul,
       volume = {9},
        pages = {1073-1085},
          doi = {10.1038/s41550-025-02571-1},
archivePrefix = {arXiv},
       eprint = {2410.02315},
 primaryClass = {astro-ph.HE},
       adsurl = {https://ui.adsabs.harvard.edu/abs/2025NatAs...9.1073S}
}

@ARTICLE{zheng2025ep240414a,
       author = {{Zheng}, Jian-He and {Zhu}, Jin-Ping and {Lu}, Wenbin and {Zhang}, Bing},
        title = "{EP240414a: Off-axis View of a Jet-cocoon System from an Expanded Progenitor Star}",
      journal = {\apj},
         year = 2025,
        month = may,
       volume = {985},
       number = {1},
          eid = {21},
        pages = {21},
          doi = {10.3847/1538-4357/adc993},
archivePrefix = {arXiv},
       eprint = {2503.24266},
 primaryClass = {astro-ph.HE},
       adsurl = {https://ui.adsabs.harvard.edu/abs/2025ApJ...985...21Z}
}

@ARTICLE{jiang2025ep240801a,
   author = {{Jiang}, S.-Q. and others},
   title = "{EP240801a/XRF 240801B: An X-ray Flash Detected by the Einstein Probe and Implications of its Multiband Afterglow}",
   journal = {ApJ},
   year = 2025,
   volume = 988,
   pages = {L34},
}

@ARTICLE{schneider2026xrf241001a,
       author = {{Schneider}, B. and {Brunet}, M. and {Gompertz}, B.~P. and {Turpin}, D. and {Malesani}, D.~B. and {Godet}, O. and {Levan}, A.~J. and {Daigne}, F. and {Sarin}, N. and {Rakotondrainibe}, N.~A. and {Martin-Carrillo}, A. and {Palmerio}, J.~T. and {Th{\"o}ne}, C.~C. and {Li}, H.~L. and {Saccardi}, A. and {de Ugarte Postigo}, A. and {Antier}, S. and {Buat}, V. and {{\v{D}}urov{\v{c}}{\'\i}kov{\'a}}, D. and {Izzo}, L. and {Leung}, J.~K. and {Mo}, G. and {Qiu}, Y.~L. and {Vergani}, S.~D. and {Wang}, J. and {Wei}, J.~Y. and {Xin}, L.~P. and {Mochkovitch}, R. and {Zhang}, B. and {Cai}, H.~B. and {Campana}, S. and {Coleiro}, A. and {Cordier}, B. and {D'Avanzo}, P. and {Dagoneau}, N. and {D'Elia}, V. and {Dong}, Y.~W. and {G{\"o}tz}, D. and {Guillot}, S. and {Han}, X.~H. and {Hartmann}, D.~H. and {Huang}, L. and {Huang}, Y.~F. and {Jakobsson}, P. and {Klotz}, A. and {Lachaud}, C. and {Le Floc'h}, E. and {Lu}, X.~M. and {Maggi}, P. and {De Pasquale}, M. and {Piron}, F. and {Salvaterra}, R. and {Schanne}, S. and {Sollerman}, J. and {Tanvir}, N.~R. and {Vidadi}, Z. and {Wang}, P. and {Wu}, C. and {Xiong}, S.~L. and {Xu}, Y. and {Zafar}, T. and {Zhang}, P.~P. and {Zhang}, S.~N. and {Zheng}, S.~J.},
        title = "{XRF 241001A/SN 2024aiiq: A faint soft X-ray transient detected by SVOM with a broad-line type Ic supernova revealed by JWST}",
      journal = {arXiv e-prints},
         year = 2026,
        month = apr,
          eid = {arXiv:2604.20346},
        pages = {arXiv:2604.20346},
          doi = {10.48550/arXiv.2604.20346},
archivePrefix = {arXiv},
       eprint = {2604.20346},
 primaryClass = {astro-ph.HE},
       adsurl = {https://ui.adsabs.harvard.edu/abs/2026arXiv260420346S}
}

@ARTICLE{srinivasaragavan2025ep250108a,
   author = {{Srinivasaragavan}, G.~P. and others},
   title = "{EP250108a/SN 2025kg: A Fast X-ray Transient with a Double-Peaked Optical Light Curve, Associated with a Broad-lined Type Ic Supernova}",
   journal = {ApJ},
   year = 2025,
   volume = 988,
   pages = {L60},
}

@ARTICLE{li2026magnetarpwn,
   author = {{Li}, Y.-F. and {Lin}, D.-B. and {Ren}, J. and {Chen}, Z.-L. and {Liang}, E.-W.},
   title = "{A Unified Model for the Emission of Supernova-Associated Fast X-ray Transients: Case Studies of EP240414a, EP250108a, and GRB 171205A}",
   journal = {ApJ},
   year = 2026,
   note = {in press},
}

@ARTICLE{chen2026ep,
       author = {{Chen}, Connery and {Wang}, Yihan and {Zhang}, Bing},
        title = "{Einstein Probe Fast X-Ray Transients Extend the Physical Parameter Space of Relativistic Jets}",
      journal = {\apjl},
         year = 2026,
        month = aug,
       volume = {1006},
       number = {2},
          eid = {L44},
        pages = {L44},
          doi = {10.3847/2041-8213/ae89a4},
archivePrefix = {arXiv},
       eprint = {2607.14317},
 primaryClass = {astro-ph.HE},
       adsurl = {https://ui.adsabs.harvard.edu/abs/2026ApJ..1006L..44C}
}

@ARTICLE{yi2025grb230307a,
       author = {{Yi}, S.-X. and {Wang}, C.-W. and {Shao}, X. and {Moradi}, R. and {Gao}, H. and {Zhang}, B. and {Xiong}, S.-L. and {Zhang}, S.-N. and {Tan}, W.-J. and {Liu}, J.-C. and {Xue}, W.-C. and {Zhang}, Y.-Q. and {Zheng}, C. and {Wang}, Y. and {Zhang}, P. and {An}, Z.-H. and {Cai}, C. and {Feng}, P.-Y. and {Gong}, K. and {Guo}, D.-Y. and {Huang}, Y. and {Li}, B. and {Li}, X.-B. and {Li}, X.-Q. and {Liu}, X.-J. and {Liu}, Y.-Q. and {Ma}, X. and {Peng}, W.-X. and {Qiao}, R. and {Song}, L.-M. and {Wang}, J. and {Wang}, P. and {Wen}, X.-Y. and {Xiao}, S. and {Xu}, Y.-B. and {Yang}, S. and {Yi}, Q.-B. and {Zhang}, D.-L. and {Zhang}, F. and {Zhang}, H.-M. and {Zhang}, J.-P. and {Zhang}, Z. and {Zhao}, X.-Y. and {Zhao}, Y. and {Zheng}, S.-J.},
        title = "{Evidence of Minijet Emission in a Large Emission Zone from a Magnetically Dominated Gamma-Ray Burst Jet}",
      journal = {\apj},
         year = 2025,
        month = jun,
       volume = {985},
       number = {2},
          eid = {239},
        pages = {239},
          doi = {10.3847/1538-4357/adcf98},
archivePrefix = {arXiv},
       eprint = {2310.07205},
 primaryClass = {astro-ph.HE},
       adsurl = {https://ui.adsabs.harvard.edu/abs/2025ApJ...985..239Y}
}

@ARTICLE{blandford1977,
       author = {{Blandford}, R.~D. and {Znajek}, R.~L.},
        title = "{Electromagnetic extraction of energy from Kerr black holes.}",
      journal = {\mnras},
         year = 1977,
        month = may,
       volume = {179},
        pages = {433-456},
          doi = {10.1093/mnras/179.3.433},
       adsurl = {https://ui.adsabs.harvard.edu/abs/1977MNRAS.179..433B}
}

@ARTICLE{tchekhovskoy2011,
   author = {{Tchekhovskoy}, A. and {Narayan}, R. and {McKinney}, J.~C.},
   title = "{Efficient generation of jets from magnetically arrested accretion on a rapidly spinning black hole}",
   journal = {MNRAS},
   year = 2011,
   volume = 418,
   pages = {L79-L83},
}

@ARTICLE{peer2004,
   author = {{Pe'er}, A. and {Waxman}, E.},
   title = "{Prompt Gamma-Ray Burst Spectra: Detailed Calculations and the Effect of Pair Production}",
   journal = {ApJ},
   year = 2004,
   volume = 613,
   pages = {448-459},
}

@ARTICLE{bromberg2016,
   author = {{Bromberg}, O. and {Tchekhovskoy}, A.},
   title = "{Relativistic MHD simulations of core-collapse GRB jets: 3D instabilities and magnetic dissipation}",
   journal = {MNRAS},
   year = 2016,
   volume = 456,
   pages = {1739-1760},
   doi = {10.1093/mnras/stv2591},
}

@ARTICLE{2024SCPMA..6789511Z,
       author = {{Zhang}, Yan-Qiu and {Xiong}, Shao-Lin and {Mao}, Ji-Rong and {Zhang}, Shuang-Nan and {Xue}, Wang-Chen and {Zheng}, Chao and {Liu}, Jia-Cong and {Zhang}, Zhen and {Wang}, Xi-Lu and {Ge}, Ming-Yu and {Yi}, Shu-Xu and {Song}, Li-Ming and {An}, Zheng-Hua and {Cai}, Ce and {Li}, Xin-Qiao and {Peng}, Wen-Xi and {Tan}, Wen-Jun and {Wang}, Chen-Wei and {Wen}, Xiang-Yang and {Wang}, Yue and {Xiao}, Shuo and {Zhang}, Fan and {Zhang}, Peng and {Zheng}, Shi-Jie},
        title = "{Observation of spectral lines in the exceptional GRB 221009A}",
      journal = {Science China Physics, Mechanics, and Astronomy},
         year = 2024,
        month = aug,
       volume = {67},
       number = {8},
          eid = {289511},
        pages = {289511},
          doi = {10.1007/s11433-023-2381-0},
archivePrefix = {arXiv},
       eprint = {2403.12851},
 primaryClass = {astro-ph.HE},
       adsurl = {https://ui.adsabs.harvard.edu/abs/2024SCPMA..6789511Z}
}

@article{zhang2024grb221009a,
  author  = {{Zhang}, Z. and {Lin}, H. and {Li}, Z. and {Xiong}, S.-L. and {Zhang}, Y.-Q. and {Zhang}, Q. and {Yi}, S.-X. and {Wang}, X.},
  title   = {{Implication of Jet Physics from MeV Line Emission of GRB 221009A}},
  journal = {The Astrophysical Journal Letters},
  year    = {2024},
  volume  = {973},
  pages   = {L17},
  doi     = {10.3847/2041-8213/ad758e},
}

@article{yi2024mevline,
  author  = {{Yi}, S.-X. and {Zhang}, Z. and {Yorgancioglu}, E.~S. and {Zhang}, S.-N. and {Xiong}, S.-L. and {Zhang}, Y.-Q.},
  title   = {{Robust constraints on the physics of the MeV emission line in GRB 221009A from optical depth arguments}},
  journal = {Monthly Notices of the Royal Astronomical Society},
  year    = {2024},
  volume  = {535},
  pages   = {982--989},
  doi     = {10.1093/mnras/stae2403}
}

@ARTICLE{jonker2013fxt,
   author = {{Jonker}, P.~G. and {Glennie}, A. and {Heida}, M. and {Maccarone}, T. and {Hodgkin}, S. and {Nelemans}, G. and {Miller-Jones}, J.~C.~A. and {Torres}, M.~A.~P. and {Fender}, R.},
   title = "{Discovery of a New Kind of Explosive X-Ray Transient near M86}",
   journal = {ApJ},
   year = 2013,
   volume = {779},
   pages = {14},
   doi = {10.1088/0004-637X/779/1/14}
}

@ARTICLE{glennie2015fxt,
   author = {{Glennie}, A. and {Jonker}, P.~G. and {Fender}, R.~P. and {Nagayama}, T. and {Pretorius}, M.~L.},
   title = "{Two fast X-ray transients in archival Chandra data}",
   journal = {MNRAS},
   year = 2015,
   volume = {450},
   pages = {3765-3770},
   doi = {10.1093/mnras/stv801}
}

@ARTICLE{quirola2022fxt,
       author = {{Quirola-V{\'a}squez}, J. and {Bauer}, F.~E. and {Jonker}, P.~G. and {Brandt}, W.~N. and {Yang}, G. and {Levan}, A.~J. and {Xue}, Y.~Q. and {Eappachen}, D. and {Zheng}, X.~C. and {Luo}, B.},
        title = "{Extragalactic fast X-ray transient candidates discovered by Chandra (2000-2014)}",
      journal = {\aap},
         year = 2022,
        month = jul,
       volume = {663},
          eid = {A168},
        pages = {A168},
          doi = {10.1051/0004-6361/202243047},
archivePrefix = {arXiv},
       eprint = {2201.07773},
 primaryClass = {astro-ph.HE},
       adsurl = {https://ui.adsabs.harvard.edu/abs/2022A&A...663A.168Q}
}

@ARTICLE{2012MNRAS.421..570G,
       author = {{Granot}, Jonathan and {Piran}, Tsvi},
        title = "{On the lateral expansion of gamma-ray burst jets}",
      journal = {\mnras},
         year = 2012,
        month = mar,
       volume = {421},
       number = {1},
        pages = {570-587},
          doi = {10.1111/j.1365-2966.2011.20335.x},
archivePrefix = {arXiv},
       eprint = {1109.6468},
 primaryClass = {astro-ph.HE},
       adsurl = {https://ui.adsabs.harvard.edu/abs/2012MNRAS.421..570G}
}

@ARTICLE{2025JHEAp..4700359Y,
       author = {{Yi}, Shu-Xu and {Yorgancioglu}, Emre Seyit and {Xiong}, S.-L. and {Zhang}, S.-N.},
        title = "{Long pulse by short central engine: Prompt emission from expanding dissipation rings in the jet front of gamma-ray bursts}",
      journal = {Journal of High Energy Astrophysics},
         year = 2025,
        month = jul,
       volume = {47},
          eid = {100359},
        pages = {100359},
          doi = {10.1016/j.jheap.2025.100359},
archivePrefix = {arXiv},
       eprint = {2411.16174},
 primaryClass = {astro-ph.HE},
       adsurl = {https://ui.adsabs.harvard.edu/abs/2025JHEAp..4700359Y}
}

@ARTICLE{2026JHEAp..5300607Y,
       author = {{Yi}, S.-X. and {Wang}, C.-W. and {Xiong}, Shao-Lin and {Zhang}, S.-N. and {Maccary}, Romain and {Moradi}, Rahim and {Xiao}, Shuo and {Feng}, Hua},
        title = "{The origin of cross-energy-similar FRED profiles in gamma-ray bursts pulses}",
      journal = {Journal of High Energy Astrophysics},
         year = 2026,
        month = jul,
       volume = {53},
          eid = {100607},
        pages = {100607},
          doi = {10.1016/j.jheap.2026.100607},
       adsurl = {https://ui.adsabs.harvard.edu/abs/2026JHEAp..5300607Y}
}

@ARTICLE{li2026ep250704a,
       author = {{Li}, An and {Wang}, Chen-Wei and {Passaleva}, Niccol{\`o} and {An}, Jie and {Zhang}, Bin-Bin and {Troja}, Eleonora and {Yin}, Yi-Han Iris and {Liu}, Yuan and {Xiong}, Shao-Lin and {Xin}, Li-Ping and {Shao}, Yi-Xuan and {Yang}, Jun and {Sun}, Hui and {Xu}, Dong and {Yang}, Yu-Han and {Ricci}, Roberto and {Gao}, He and {Antier}, Sarah and {Becerra}, Rosa L. and {Cao}, Jia-Xin and {Castro-Tirado}, Alberto Javier and {Chen}, Xin-Lei and {Cheng}, Ye-Hao and {Chen}, Yong and {Cheng}, Hua-Qing and {D'Elia}, Valerio and {De Pasquale}, Massimiliano and {Dong}, Yong-Wei and {Elhosseiny}, Eslam and {Eyles-Ferris}, Rob A.~J. and {Gritsevich}, Maria and {Han}, Xu-Hui and {Hartmann}, Dieter and {Hu}, You-Dong and {Hu}, Jing-Wei and {Jia}, Shu-Mei and {Kochiashvili}, Nino and {Lei}, Wei-Hua and {Levan}, Andrew J. and {Li}, Cheng-Kui and {Li}, Dong-Yue and {Li}, Hua-Li and {Li}, Xiao-Bo and {Ling}, Zhi-Xing and {Liu}, He-Yang and {Lv}, Hou-Jun and {Malesani}, Daniele B. and {O'Connor}, Brendan and {Pan}, Hai-Wu and {Bhushan Pandey}, Shashi and {Perez-Garcia}, Ignacio and {Pieterse}, Dani{\"e}lle L.~A. and {Pillas}, Marion and {Qiu}, Yu-Lei and {Saccardi}, Andrea and {S{\'a}nchez-Ram{\'\i}rez}, Rub{\'e}n and {Tan}, Wen-Jun and {Tanasan}, Manasanun and {Tanvir}, Nial R. and {Vergani}, Susanna D. and {Wang}, Jing and {Wang}, Xiao-Feng and {Wu}, Qin-Yu and {Yi}, Shu-Xu and {Yusufjon}, Tillayev and {Zhang}, Chen and {Zhang}, Wen-Da and {Zhang}, Yi-Jia and {Zhao}, Guo-Ying and {Zheng}, Chao and {Zheng}, Shi-Jie and {Zhou}, Chang and {Zhou}, Ping and {Cordier}, Bertrand and {Wei}, Jian-Yan and {Yuan}, Weimin and {Zhang}, Shuang-Nan and {Zhang}, Bing},
        title = "{Minutes-long soft X-ray prompt emission from a compact object merger}",
      journal = {arXiv e-prints},
         year = 2026,
        month = jan,
          eid = {arXiv:2601.14137},
        pages = {arXiv:2601.14137},
          doi = {10.48550/arXiv.2601.14137},
archivePrefix = {arXiv},
       eprint = {2601.14137},
 primaryClass = {astro-ph.HE},
       adsurl = {https://ui.adsabs.harvard.edu/abs/2026arXiv260114137L}
}

@ARTICLE{sun2025grb230307a,
       author = {{Sun}, H. and {Wang}, C.-W. and {Yang}, J. and {Zhang}, B.-B. and {Xiong}, S.-L. and {Yin}, Y.-H.~I. and {Liu}, Y. and {Li}, Y. and {Xue}, W.-C. and {Yan}, Z. and {Zhang}, C. and {Tan}, W.-J. and {Pan}, H.-W. and {Liu}, J.-C. and {Cheng}, H.-Q. and {Zhang}, Y.-Q. and {Hu}, J.-W. and {Zheng}, C. and {An}, Z.-H. and {Cai}, C. and {Cai}, Z.-M. and {Hu}, L. and {Jin}, C. and {Li}, D.-Y. and {Li}, X.-Q. and {Liu}, H.-Y. and {Liu}, M. and {Peng}, W.-X. and {Song}, L.-M. and {Sun}, S.-L. and {Sun}, X.-J. and {Wang}, X.-L. and {Wen}, X.-Y. and {Xiao}, S. and {Yi}, S.-X. and {Zhang}, F. and {Zhang}, W.-D. and {Zhang}, X.-F. and {Zhang}, Y.-H. and {Zhao}, D.-H. and {Zheng}, S.-J. and {Ling}, Z.-X. and {Zhang}, S.-N. and {Yuan}, W. and {Zhang}, B.},
        title = "{Magnetar emergence in a peculiar gamma-ray burst from a compact star merger}",
      journal = {National Science Review},
         year = 2025,
        month = mar,
       volume = {12},
       number = {3},
          eid = {nwae401},
        pages = {nwae401},
          doi = {10.1093/nsr/nwae401},
archivePrefix = {arXiv},
       eprint = {2307.05689},
 primaryClass = {astro-ph.HE},
       adsurl = {https://ui.adsabs.harvard.edu/abs/2025NSRev..12E.401S}
}

@ARTICLE{ruffini1999pair,
   author = {{Ruffini}, R. and {Salmonson}, J.~D. and {Wilson}, J.~R. and {Xue}, S.-S.},
   title = "{On the pair electromagnetic pulse of a black hole with electromagnetic structure}",
   journal = {A\&A},
   year = 1999,
   volume = 350,
   pages = {334-343},
   doi = {10.48550/arXiv.astro-ph/9907030},
   eprint = {astro-ph/9907030},
   archivePrefix = {arXiv},
}

@ARTICLE{ruffini2000pair,
   author = {{Ruffini}, R. and {Salmonson}, J.~D. and {Wilson}, J.~R. and {Xue}, S.-S.},
   title = "{On the pair-electromagnetic pulse from an electromagnetic black hole surrounded by a baryonic remnant}",
   journal = {A\&A},
   year = 2000,
   volume = 359,
   pages = {855-864},
   doi = {10.48550/arXiv.astro-ph/0004257},
   eprint = {astro-ph/0004257},
   archivePrefix = {arXiv},
}

@ARTICLE{bianco2001spike,
   author = {{Bianco}, C.~L. and {Ruffini}, R. and {Xue}, S.-S.},
   title = "{The elementary spike produced by a pure e+e- pair-electromagnetic pulse from a black hole: the PEM Pulse}",
   journal = {A\&A},
   year = 2001,
   volume = 368,
   pages = {377-390},
   doi = {10.1051/0004-6361:20000556},
   eprint = {astro-ph/0102060},
   archivePrefix = {arXiv},
}

@ARTICLE{giannios2012,
   author = {{Giannios}, D.},
   title = "{The peak energy of dissipative gamma-ray burst photospheres}",
   journal = {MNRAS},
   year = 2012,
   volume = 422,
   pages = {3092-3098},
   doi = {10.1111/j.1365-2966.2012.20845.x},
}

@ARTICLE{beloborodov2013,
   author = {{Beloborodov}, A.~M.},
   title = "{Regulation of the Spectral Peak in Gamma-Ray Bursts}",
   journal = {ApJ},
   year = 2013,
   volume = 764,
   pages = {157},
   doi = {10.1088/0004-637X/764/2/157},
}

@ARTICLE{beloborodov2017,
   author = {{Beloborodov}, A.~M. and {M{\'e}sz{\'a}ros}, P.},
   title = "{Photospheric Emission of Gamma-Ray Bursts}",
   journal = {Space Science Reviews},
   year = 2017,
   volume = 207,
   pages = {87-110},
   doi = {10.1007/s11214-017-0377-1},
}

@ARTICLE{samuelsson2023,
   author = {{Samuelsson}, F. and {Ryde}, F.},
   title = "{Observational Characteristics of Radiation-mediated Shocks in Photospheric Gamma-Ray Burst Emission}",
   journal = {ApJ},
   year = 2023,
   volume = 956,
   pages = {42},
   doi = {10.3847/1538-4357/acf564},
}

@ARTICLE{pozdnyakov1983,
   author = {{Pozdnyakov}, L.~A. and {Sobol}, I.~M. and {Syunyaev}, R.~A.},
   title = "{Comptonization and the shaping of X-ray source spectra - Monte Carlo calculations}",
   journal = {Astrophysics and Space Physics Reviews},
   year = 1983,
   volume = 2,
   pages = {189},
}

@ARTICLE{kennel1984,
       author = {{Kennel}, C.~F. and {Coroniti}, F.~V.},
        title = "{Confinement of the Crab pulsar's wind by its supernova remnant.}",
      journal = {\apj},
         year = 1984,
        month = aug,
       volume = {283},
        pages = {694-709},
          doi = {10.1086/162356},
       adsurl = {https://ui.adsabs.harvard.edu/abs/1984ApJ...283..694K}
}

@ARTICLE{zhang2005,
       author = {{Zhang}, Bing and {Kobayashi}, Shiho},
        title = "{Gamma-Ray Burst Early Afterglows: Reverse Shock Emission from an Arbitrarily Magnetized Ejecta}",
      journal = {\apj},
         year = 2005,
        month = jul,
       volume = {628},
       number = {1},
        pages = {315-334},
          doi = {10.1086/429787},
archivePrefix = {arXiv},
       eprint = {astro-ph/0404140},
 primaryClass = {astro-ph},
       adsurl = {https://ui.adsabs.harvard.edu/abs/2005ApJ...628..315Z}
}

@ARTICLE{giannios2008,
       author = {{Giannios}, D. and {Mimica}, P. and {Aloy}, M.~A.},
        title = "{On the existence of a reverse shock in magnetized gamma-ray burst ejecta}",
      journal = {\aap},
         year = 2008,
        month = feb,
       volume = {478},
       number = {3},
        pages = {747-753},
          doi = {10.1051/0004-6361:20078931},
archivePrefix = {arXiv},
       eprint = {0711.1980},
 primaryClass = {astro-ph},
       adsurl = {https://ui.adsabs.harvard.edu/abs/2008A&A...478..747G}
}

@ARTICLE{mimica2010,
       author = {{Mimica}, P. and {Giannios}, D. and {Aloy}, M.~A.},
        title = "{Multiwavelength afterglow light curves from magnetized gamma-ray burst flows}",
      journal = {\mnras},
         year = 2010,
        month = oct,
       volume = {407},
       number = {4},
        pages = {2501-2510},
          doi = {10.1111/j.1365-2966.2010.17071.x},
archivePrefix = {arXiv},
       eprint = {1004.2720},
 primaryClass = {astro-ph.HE},
       adsurl = {https://ui.adsabs.harvard.edu/abs/2010MNRAS.407.2501M}
}
\end{document}